\documentclass[prl,twocolumn,amsmath,amssymb,nofootinbib]{revtex4}
\usepackage{amsmath}
\usepackage{amssymb}
\usepackage{amsfonts}
\usepackage{eucal}

\newcommand{\be}{\begin{equation}}
\newcommand{\bse}{\begin{subequations}}
\newcommand{\ese}{\end{subequations}}
\newcommand{\bea}{\begin{eqnarray}}
\newcommand{\eea}{\end{eqnarray}}
\newcommand{\ba}{\begin{array}}
\newcommand{\ea}{\end{array}}
\newcommand{\ee}{\end{equation}}

\def\Pl{{\rm Pl}}
\def\hp{h_+}
\def\hc{h_{\times}}
\def\half{\frac{1}{2}}

\def\yboxit#1#2{\vbox{\hrule height #1 \hbox{\vrule width #1
    \vbox{#2}\vrule width #1 }\hrule height #1 }}
\def\fillbox#1{\hbox to #1{\vbox to #1{\vfil}\hfil}}
\def\ybox{\yboxit{0.4pt}{\fillbox{8pt}}\hskip-0.4pt}
\def\VEV#1{\langle{ #1} \rangle}

\begin{document}

\title{Is Cosmic Parity Violation Responsible for the Anomalies in the WMAP Data?}
\author{Stephon H.S Alexander}  
\affiliation{Depts. of Physics and Astronomy, Institute for Gravitational Physics and Geometry, Center For Gravitational Wave Physics\\
The Pennsylvania State University, 104 Davey Lab, University Park, PA 16802} 

\begin{abstract}
This research argues that parity violation in general relativity can simultaneously explain the observed loss in power and alignment at a preferred axis ('Axis of Evil')in the low multipole moments of the WMAP data.  This observational possibility also provides an experimental window for an inflationary leptogenesis mechanism arising from large-scale parity violation.   A velocity dependent potential is induced from gravitational backreaction, which modifies the primordial scalar angular power spectrum of density perturbations in the CMB.  This modification suppresses power of odd parity multipoles on large scales which can be associated with the scale of a massive right-handed neutrino.   \end{abstract}
\maketitle
\section{Introduction}

Parity (P) violation is one of the cornerstones of the current standard model of particle interactions.  Nonetheless, this violation only occurs within the weak interaction sector.   It is still unknown how parity violation arises from a unified scheme which includes the other forces, in particular, gravity.   Over the past two decades, a wealth of cosmological data has confirmed the theoretical predictions of the inflationary paradigm\cite{alan,WMAP}.  Inflation posits that specific initial conditions encoded in quantum fluctuations are responsible for large scale structure in the universe.  Combining these ideas, it is plausible that the initial conditions set by inflation are parity violating.  It is interesting to ask what the observational consequences will be. 

If parity violation on large scales can coexist with a homogeneous and isotropic universe like ours how can we observe it?  This question has been asked in the past and analyzed in the context of the CMB polarization \cite{LWK,Balaji:2003sw}.    It was found that the direct signal would be undetectable in the most optimistic cases\cite{LWK}.  The authors of \cite{APS} found that parity violation sourced by a non-vanishing phase of a pseudo-scalar inflaton field can provide all Sakharov conditions for leptogenesis.  While this mechanism is compelling, it is still necessary to seek its observational consequences, especially in recent and future CMB observations.    

In principle, the parity violation in general relativity leads to leptogenesis by transmitting itself into B-L violation through primordial gravitational waves\cite{APS,Alexander:2004wk,Alexander:2004xd} .  This occurs because there is a gravitational Chern-Simons coupling to a pseudo-scalar field which is generated through a Green-Schwarz mechanism\cite{GS}.  The Chern-Simons operator in the inflationary background is non-vanishing.    A key point of this letter is that the Chern-Simons operator gives a contribution to the Energy-Momentum tensor leading to a suppression of the odd-parity modes in the powerspectrum.
  
In this letter I will show that if parity is violated  during the inflationary epoch, that only the large scale, odd-parity, perturbations of the inflation field will experience a loss of power.  This happens because   gravitational backreaction induces a velocity dependent potential for the primordial scalar fluctuations.  At the same time the gravitational backreaction will produce leptons.  The power suppression will cease for large multipoles, which coincides with energy scales comparable to a massive right-handed neutrino.   I will explore the degree to which large-scale parity violation and, hence, leptogenesis can be observable in the WMAP data.  Another goal of this paper is to identify the origin of the two persistent anomalies in the WMAP data, the loss of power on large scales and an alignment of multipoles along a preferred axis for low $\ell$'s, (the so called ``Axis of Evil'') which has been argued to suppress odd parity modes at  $\ell=2-5$\cite{Land:2005ad,Land:2005jq,copi,copiii}.  

\section{General Relativity and Parity Violation}
General Relativity can readily be extended to have parity violation by including a Chern-Simons term.  For homogenous and isotropic space-times, such as de Sitter and FRW, this term vanishes.  But in the presence of a rolling pseudo scalar field this is no longer the case.   I will demonstrate this fact and its consequences on the scalar power-spectrum.  First, let us consider the extended Gravity action with a Chern-Simons extension\cite{Ashtekar:1988sw}.

\be \label{action} S = \int d^{4}x\sqrt{-g} (R -\frac{1}{2} \partial_{\mu} \phi \partial^{\mu}\phi + V(\phi) + {\phi\over M_{Pl}} \epsilon^{\alpha\beta\gamma\delta}
 R_{\alpha\beta \rho\sigma} R_{\gamma\delta}{}^{\rho\sigma} ) \ee
 integrating the last term in the above equation by parts yields its relation to the three dimensional Chern-Simons term.

\be \int\phi R\wedge R =-\int d\phi \wedge \Omega_{CS} \ee
where $\Omega_{CS} = d\omega\wedge \omega +\frac{2}{3}\omega\wedge\omega\wedge\omega $.
The general form of metric perturbations about an FRW universe can be 
parameterized as
\be
\begin{split}
ds^2 &= -(1+2\varphi)dt^2+w_i dtdx^i\cr
&+a^2(t)\left[\left((1+2\psi)\delta_{ij}+h_{ij}\right)dx^idx^j\right]
\end{split}
\ee
where $\varphi$, $\psi$, $w_i$ and $h_{ij}$ respectively parametrize the
 scalar, vector, and tensor fluctuations of the metric.  It is 
straightforward to 
show that the scalar and vector perturbations do not contribute to 
$R\tilde R$, and so these fluctuations are ignored in the following discussion.
We can also fix a gauge so that the tensor fluctuation is parametrized by the
two physical transverse traceless elements of $h_{ij}$.  For gravity
waves moving in the $z$ direction without loss of generality, one writes
\bea\label{mytensors}
ds^2&=&-dt^2+a^2(t)\bigl[(1-\hp)dx^2\cr
&+&(1+\hp)dy^2 + 2\hc dxdy +dz^2\bigr]
\eea
where $a(t)  = e^{Ht}$ during inflation and $\hp$, $\hc$ are functions of
$t$, $z$.
To see the CP violation more explicitly, it is convenient to use a 
helicity basis
\be
h_L =  (\hp - i \hc){/\sqrt{2}} \ , \qquad 
h_R = (\hp + i \hc){/\sqrt{2}} \ .
\ee
Here $h_L$ and $h_R$ are complex conjugate scalar fields. To be very explicit,
the negative frequency part of $h_L$ is the conjugate of the positive 
frequency part of $h_R$, and both are built from wavefunctions for left-handed
gravitons.

The contribution of tensor perturbations to $R\tilde R$, up to 
second order in $h_L$ and $h_R$, is
\be\label{RRdual} 
\begin{split}
R\tilde R= \frac{4i}{a^{3}}\biggl[&\bigg(
{\partial^2_z} h_R\ {\partial_{z}\partial_{t}}h_L+a^2 {\partial^2_{t}}h_R\ 
{\partial_{t}\partial_{z}}h_L\cr
& +\frac{1}{2}{\partial_t}a^2 {\partial_t} h_R\ 
{\partial_{t}\partial_{z}}h_L\bigg)
-{\left(L\leftrightarrow R\right)}
\biggr]
\end{split}
\ee
If $h_L$ and $h_R$ have the same dispersion relation, this expression
vanishes. Thus, for $R\tilde R$  to 
be nonzero, one needs a 
 `cosmological birefringence' during inflation.  Such an effect is induced 
by the addition of the pseudo-scalar Chern-Simons coupling to the gravitational 
equations~\cite{LWK}.   In the presence of this source the gravitational waves obey the following equation.

\be\label{LReqs}
  \ybox\, h_L = - 2i {\Theta\over a} {\dot h}^\prime_L \ , \qquad 
  \ybox\, h_R = + 2i {\Theta\over a} {\dot h}^\prime_R \ ,  
\ee
where
\be\label{Thetaval}
  \Theta = 8 (\frac{H}{M_\Pl})^{2} \dot\phi)/HM_\Pl \ ,
\ee
dots denote time dervatives, and primes denote differentiation of $F$
with respect to $\phi$.

The authors of \cite{APS} found exact solutions to the above equations and evaluated the Green's functions.  A one loop effect ensures a VEV for \ref{RRdual} and in the presence of the above tensor perturbations this term will not vanish, provided that the pseudo-scalar field which couples to it is dynamical.   Let us study the effects of this term on the scalar field evolution.  The equation of motion for the pseudo-scalar is:

\be \label{phi} \ddot{\phi} + 3H\dot{\phi} +k^{2}\phi + \frac{dV}{d\phi} = -<R\tilde{R}> \ee

The VEV on the RHS can be calculated from evaluating the Greens function of the gravity waves.  This was previously done by \cite{APS}; the answer is:
\be\label{RRdualval}
  \VEV{R\tilde R} =  {16\over a}\, \int \, {d^3 k\over (2\pi)^3}\ 
  {H^2\over 2 k^3 M_\Pl^2} 
                      (k\eta)^2 \cdot k^4 \Theta )
\ee

Plugging this result into eq(\ref{phi}) yields,
\be \label{phitwo} \ddot{\phi} + 3H\dot{\phi} +k^{2}\phi + \frac{dV}{d\phi} =   {16\over a}\, \int \, {d^3 k\over (2\pi)^3}\ 
  {H^2\over 2 k^3 M_\Pl^2} 
                      (k\eta)^2 \cdot k^4 \Theta \ee
                      
 One would like to calculate the power-spectrum of the scalar field, but the source makes the equation non-linear.   Heuristically, the $k$-dependent source can lead to enhancement/suppression at a characteristic wavenumber.  In what follows this problem is resolved by showing that the Chern-Simons coupling can modify the Energy-Momentum tensor leading to a modified Friedmann equation.  It is through this route that one will be able to calculate the modified power-spectrum.                 
                      
                      



\section{Modified Energy Momentum Tensor}
I will now show that the Chern-Simons term induces a modification to the Energy-Momentum tensor.  This will give us a velocity dependent contribution to the Energy Momentum tensor which will modify the power-spectrum in a specific way to be discussed in the next section.
Variation of the Chern-Simons action with respect to the metric yield the correction to the Energy-Momentum tensor\cite{roman}.

\be \delta I_{CS} = \delta \int d^{4}x\phi ^{*}RR=\int d^{4}x\sqrt{-g}C_{\mu\nu}\delta g^{\mu\nu} \ee
where $^{*}R_{\mu\nu} =\epsilon_{\alpha\beta\mu\nu}R^{\alpha\beta}$ and
\be
\begin{split} 
 C^{\mu\nu} =\frac{1}{-2\sqrt{-g}}[v_{\sigma}(\epsilon^{\sigma\mu\alpha\beta}D_{\alpha}R^{\nu}_{\beta} + \epsilon^{\sigma\nu\alpha\beta}D_{\alpha}R^{\alpha}_{\beta})\cr
 + v_{\sigma \tau}(^{*}R_{\tau\mu\sigma\nu} + ^{*}R^{\tau\nu\sigma\mu})] 
\end{split} 
\ee

where $C_{\mu\nu}$ is the Cotton tensor,  $v_{\sigma\tau}$ is the covariant derivative of the embedding coordinate and in terms of the pseudo-scalar field $v_{\sigma\tau} = D_{\sigma}D_{\tau} \phi$ and $v_{\sigma}=D_{\sigma}\phi$.  Therefore, Einstein's equation is modified in the following way.

\be G^{\mu\nu} = -(8\pi GT^{\mu\nu} 
+ C^{\mu\nu}) \ee

The covariant divergence of the Cotton tensor is non-zero.

\be \label{EM} D_{\mu}C^{\mu\nu} = \frac{1}{8\sqrt{-g}}\partial^{\nu}\phi ^{*}R R \ee

One can now derive the contribution to the Energy-Momentum tensor $T^{00}$ by integrating \ref{EM} over time.  This is possible in a purely homogenous and isotropic space-time where the pseudo-scalar field is only time dependent (i.e. $\partial_{\mu}\phi= \dot{\phi}$).  We obtain.

\be T_{CS}^{00} =\frac{1}{HM_{pl}}( \dot{\phi}^{*}R R - e^{-tH}) \ee

With the energy momentum tensor at hand the effective Lagrangian for the pseudo-scalar field becomes:
\be \label{dual} \cal{L}\rm_{\phi} = -\partial_{\mu}\phi \partial^{\mu} \phi + V(\phi) + \frac{1}{HM_{pl}} \dot{\phi}*R R \ee

Plugging in the expression for the Chern-Simons term (\ref{RRdual}) gives, 

\be \cal{L}\rm_{\phi} = -\partial_{\mu}\phi \partial^{\mu} \phi + V(\phi) + \dot{\phi}^{2}{ f(t(k))\over H M_{Pl}}   \ee
from eq \eqref{RRdualval},
\be f(t(k)) ={\cal{N}\rm\over a(t)}  \int \, {d^3 k\over (2\pi)^3 HM_{\Pl}}\ 
       {1\over 16\pi^2}\, {8 H^4  k^3 \eta^2  \over M_\Pl^4}  \ 
       .\ee
where $\dot{\phi} f(t(k)) = <R\wedge R>$.  After performing the momentum integral,
\be f(t(k)) ={\cal{N}\rm \over(2\pi)^{3} }a^{-3} (\frac{H}{M_{Pl}})^{4}(\frac{k_{s}}{M_{Pl}})({k_{s}\over H
})^{3} k^{2}_{s}, \ee
where $k_{s}$ is the UV cutoff in the momentum integral, which in this case is identified with the scale of the right handed neutrino.

The gravitational backreaction from the coupling of the Chern-Simons term and the pseudo-scalar induces a velocity dependent potential.  This is the key to understanding how the power-spectrum can be modified from space-time parity violation.  

\section{Modified Power Spectrum and Parity Violation}

 Due to arguments in the previous  section the effective action for the inflaton, neglecting spatial gradients,  gets modified to the following form:
\be S= \int d^{4}x \sqrt{-g} \big(V(\phi) -\frac{1}{2}(1 - f(t(k)))\dot{\phi}^{2} \big) \ee

The leptogenesis mechanism discussed by \cite{APS} relied on the well-known fact that
the lepton number current, and
also the total fermion number current, has a gravitational anomaly in the 
Standard Model.  Explicitly,
\be\label{Jlepton}
     \partial_\mu J^\mu_\ell  =  {3\over 16\pi^2}   R  \tilde R
\ee
where
\be
J^\mu_\ell =    \bar \ell_i\gamma^\mu \ell_i + \bar \nu_i 
\gamma^\mu \nu_i \  ,\ 
R\tilde R = \half \epsilon^{\alpha\beta\gamma\delta}
 R_{\alpha\beta \rho\sigma} R_{\gamma\delta}{}^{\rho\sigma} \ .
\ee
The anomaly requires an imbalance of left- and right-handed leptons.  In general, \eqref{Jlepton} will be 
correct in an effective theory valid below a scale $\mu$.  Therefore, a non-vanishing $f$ signifies lepton number production.   It is interesting that leptogenesis is directly conneted to the modification of the power-spectrum during inflation.  

Another important fact is that there will be an additional contribution to the velocity dependent potential which cancels $f(k)$ when the momentum of the gravity waves approaches the scale $\mu$ of the right handed neutrino.

\be f(k,t) = f_{l}(k=\mu,t) - f_{r}(k=\mu,t) \rightarrow 0 \ee
where $f_{l}$ and $f_{r}$ are the left and right handed contribution to the gravitational anomaly integral, respectively.
I will assume that  this scale corresonds to the scale of comoving wavenumber 60 e-foldings before inflation ends.  

Given the above modification, one can calculate the primordial power spectrum $P_{H}(k(t_{f}))$ for modes with comoving wavenumber $k(t_{f})$ which re-enter the horizon given the primordial power spectrum for modes with comoving wavenumber $k(t_{i})$ which exit the horizon at time $t_{i}$ with the equation.
\be \label{power} P(k,t_{f})_{H} = \frac{1 + \omega(t(k)_{f})}{1 + \omega(t(k)_{i})} P(k,t_{i})_{H}^{0} \ee
where the equation of state will be modified due to the correction of the kinetic energy.
We can compute the modified powerspectrum by using the following identities:

\be \omega = \frac{p}{\rho}\ee
and
\be p = \frac{1}{2}\dot{\phi}^{2} + V \ee
\be \rho = \frac{1}{2}\dot{\phi}^{2} - V,\ee

After a little algebra, we obtain:
 
\be \label{spectrum} P(k)_{H} = \frac{P(k)^{0}}{1+ \frac{f(t(k))}{H}} \ee

with the above modification for the primordial power-spectum, the physical mechanism for power suppression becomes clear.  Using [\ref{spectrum}] leads to a modified powespectrum of the following form:
\be  P(k)_{H} = \frac{P(k)_{H}^{0}}{(1 + {\cal{N}\rm \over(2\pi)^{3} }a^{-3} (\frac{H}{M_{Pl}})^{4}(\frac{k_{s}}{M_{Pl}})^{2}({k_{s}\over H
})^{4} )}\ee

The modified power-spectrum has the following interesting profile.  Since $\ell$  is proportional to the co-moving wavenumber $k$, the power-spectrum for all $\ell$ larger than the scale of the right-handed neutrino($k_{s}$) will also remain unmodified because $f(k \geq k_{s})$ vanishes due to anomaly cancellation.  This means that the $C_{\ell}$'s for large $\ell$'s in the WMAP data will not be affected.

 However, the function $f(t,k<k_{s})$ will not vanish for all $l$'s less than this characteristic scale.  One can estimate this number by setting $k_s=10^{14} GeV$, a sensible upper-limit for the scale of the right handed neutrino.  A sensible value for $\cal{N}\rm$ is $10^{3}$. This value yields:
 
\be {f(t(k))\over H M_{Pl}}  \sim 10^{-22} \times e^{N(k)} \ee
and the power-spectrum becomes,
\be  P(k)_{H} = \frac{P_{H}^{0}}{(1 + 10^{-22}\times e^{3N(k)} )}.\ee
where $e^{N(k)} = \frac{a(t)_{0}}{a(t)}.$
The correction exponentially depends on the number of e-foldings.  If we assume that the scale of the right handed neutrino corrseponds to $l=2$ (60 -efoldings before the end of inflation) then the power is almost completely suppressed.  Moreover, since anomaly cancellation occurs when the right handed neutrino scale is reached, the power will only be suppressed at low multipole moments less than the scale of the right-handed neutrino.  Hence, the scale of the right-handed neutrino can be identified as the turning point in the WMAP data wherein the power spectrum deviates from standard inflationary predictions.  It is interesting that the source which suppresses the power  arises from the most general parity violating operator in general relativity. 
\section{Power Suppression and the Axis of Evil}
Recently, Land and Magueijo found a preferred frame in the WMAP data which is significantly aligned for multipoles $l=2,3,4,5$ which defines an overall preferred axis\cite{Land:2005jq}.  For these multipoles, the $m's$ turn out to be preferred such that $l+m= even$.  In other words, multipoles are dominated by positive mirror parity in this preferred frame.  To identify the culprit of the 'Axis of Evil' requires an understanding of why odd parity modes in this frame are supressed at  low multipole moments.  I now argue that the gravitational parity violation via. the Chern-Simon's term can be the culprit behind the 'Axis of Evil'

Consider the equation of motion of the inflaton field in our model.
\be \ddot{\phi} + 3H\dot{\phi} + \nabla^{2}\phi =  <R\wedge R> .\ee
 
The Chern-Simon's term on the R.H.S has odd parity.  This is the key to understanding the observation of even mirror-parity in the WMAP data.  To see this, let us return to the modified powerspectrum, \eqref{spectrum}
 
\be P(k)_{H} = \frac{P(k)_{H}^{0}}{(1+ \frac{f(t(k))}{ H M_{Pl}})} \ee
we see from \eqref{dual} that the function $f(t(k))$ which suppreses the power is also proportional to the odd parity Chern-Simon's term.  Therefore, only the odd parity modes will be supressed in this mechanism.  Conversley, the even parity mode of the power-spectrum will not be modified in the Axis of Evil. This is consistent with the preferred $m's$ which are not supressed such that parity even modes survive.  According to our mechanism, $l+m=odd$ modes are supressed as observed in the Axis of Evil effect.   We will pursue this issue in a forthcoming paper \cite{SM}.

\section{conclusion}
In this letter, I demonstrate that parity violation and leptogenesis in the early universe can tie together the two persistent anomalies in the CMB; loss of power and the alignment of low multipole moments along a preferred axis which has even mirror parity; the so called 'Axis of Evil'.  We see this mechanism as a first step toward identifing a possible physical mechanism for the alignment anomalies.  However, this mechanism does not address choice of the plane where even-mirror parity persists.  It is plausable that this choice is connected to a special three-topology which leads to a non-vanishing Chern-Simons number, such as those described by Witt \cite{Morrow-Jones:1993zu}.

In particular, I have shown that when general relativity is extended to have large-scale parity violation during the inflationary epoch, by the inclusion of a gravitational Chern-Simons term, the backreaction of gravitational waves  can lead to suppression of the odd parity modes in the primordial scalar power-spectrum at scales less than the right-handed neutrino.   Physically this loss of power signifies lepton number production during the inflationary epoch.  This is clear since the Chern-Simons term in the inflationary background simultaneously modifies the Energy-Momentum tensor of the scalar field and is proportional to the rate of lepton number density.  In a future work, \cite{Sam} we will report on a more systematic treatment of the data in light of the leptgenesis mechanism of \cite{APS}.
\begin{acknowledgments}
The  author gives special thanks to Andrew Liddle, Kate Land, Subodh Patil and Peter Sheppard for their constructive criticisms and discussions.  I also thank Robert Brandenberger, Shahin Sheikh-Jabbari, Lee Smolin and Michael Peskin for discussions. 
\end{acknowledgments}

\end{document}